\documentstyle[aps,prl,eqsecnum,preprint,tighten]{revtex}

\begin{document}
\def\bq{\begin{equation}}
\def\eq{\end{equation}}

\title
{
UNIVERSAL SINGULARITY AT THE CLOSURE OF A GAP IN A RANDOM MATRIX THEORY}
\author
{ E. Br\'ezin$^{1}$ and S. Hikami$^{2}$} 
\address
{$^{1}$ Laboratoire de Physique Th\'eorique, Ecole Normale
Sup\'erieure, 
24 rue Lhomond 75231, Paris Cedex 05, France{\footnote{
Unit\'e propre du centre national de la Recherche Scientifique, Associ\'ee
\`a l'Ecole Normale Sup\'erieure et \`a l'Universit\'e de Paris-Sud} 
}\\
$^{2}$ Department of Pure and Applied Sciences, University of Tokyo\\
{Meguro-ku, Komaba, Tokyo 153, Japan}\\
}
\maketitle
\vskip 3mm
\begin{abstract}

We consider a Hamiltonian $ H = H_0+ V $, in which $ H_0$ is a given
non-random Hermitian matrix,
and
$V$ is an $N \times N$ Hermitian random matrix with a Gaussian probability
distribution.
We had shown before that Dyson's  universality of the short-range correlations
between energy levels holds at generic points of the spectrum independently
 of  $H_{0}$. We consider here the case in which the spectrum of  $H_{0}$
is such that there is
a gap in the average density of eigenvalues of $H$ which is thus split into
two pieces.
When the spectrum of $H_{0}$ is tuned so that the gap closes, a new class
of universality
appears for the energy correlations in the vicinity of this singular point.

\end{abstract}
\newpage
\section{INTRODUCTION}
\vskip 5mm

We consider a Hamiltonian which is the sum of a given deterministic part
$H_0$ and of a random
{\it}potential $V$ with a Gaussian probability distribution.
Although the measure is not unitary invariant,  one can still obtain the
probability distribution
for the eigenvalues  of $H$ through the well-known
Itzykson-Zuber integral \cite{Itzykson}. Generalizing a method introduced
 by Kazakov \cite{Kazakov} for the density of eigenvalues,
we have found an exact
representation of the n-level correlation functions in terms of the
determinant of
an $n\times n$ matrix whose matrix elements are given by a kernel
$K_N(\lambda,\mu)$ .
\cite{BH1,BH2,BH3}. For generic values of $\lambda$ and $\mu$ in the
support of the average
spectrum,  we proved earlier that this kernel reduces universally to  the
sine-kernel of
Dyson
\cite{Dyson} as if $H_0$ was not there. Consequently all the correlation
functions,including the
level-spacing distribution,  reduce to the Wigner-Dyson form in the short
distance regime
independently of $H_{0}$.
However, at singular points of the spectrum the situation is different. For
instance at the edge of
the spectrum  of the density of state, the kernel is given in terms of Airy
functions instead and a
new class of universality for the correlations appears
\cite{Brezin,Tracey,BHZ2}.
In this paper, we investigate what happens when  two edge-singularities
collapse. The spectrum of
$H_0$ is thus tuned to produce a gap in the average density of eigenvalues
of $H$, which closes at
the origin through a fine tuning of the parameters. The simplest way is to
take $\pm 1$ for the
eigenvalues of $H_0$, with an equal number of positive and negative
eigenvalues, but we shall prove
that the results are independent of $H_0$, provided a gap closes. Near the
origin, a new
universal behavior appears, which is not of Airy type. The kernel which
governs this new singularity
will be discussed in detail. A relation to  Painlev\'e II differential
equations and to a
two-dimensional $A_2$ Garnier
 system is found. Higher multi-critical behavior is also investigated.

We also consider the analogous problem with $V$ made of complex blocks, and
in that case a similar tuning of  $H_0$ leads to the degeneracy of
a Bessel kernel, related to Painlev\'e  III equations.

\section{ DETERMINISTIC PLUS RANDOM HAMILTONIAN}
\vskip 5mm

We consider an $N \times N$ Hamiltonian matrix $H = H_0 + V$, where $H_0$
is a given,
non-random Hermitian matrix, and
$V$ is a random Gaussian
Hermitian matrix.
The probability distribution $P(H)$ is thus given by
\begin{eqnarray}\label{2.1}
P(H) &=& {1\over{Z}}e^{ - {N\over{2}} {\rm Tr} V^2 }\nonumber\\ &=&
{1\over{Z^{'}}}e^{- {N\over{2}}{\rm Tr} ( H^{2}- 2 H_0 H)} \end{eqnarray}

We are thus dealing with a Gaussian unitary ensemble modified by the
external matrix source
$H_0$, which breaks the unitary invariance of the measure. In previous work
\cite{BH1,BH2,BH3},
 we have discussed the density of state,
 the two-level correlation function and the n-level correlations. For
completeness we briefly
recall here a few steps, but refer the interested reader to our earlier work.
The density of state $\rho(\lambda)$ is
\begin{eqnarray}\label{2.2}
\rho(\lambda) &=& {1\over{N}} < {\rm Tr} \delta ( \lambda - H ) > \nonumber\\
&=&
\int_{-\infty}^{+\infty} {dt\over{2 \pi}} e^{- iNt \lambda} U(t) \end{eqnarray}
where $U(t)$ is the average "evolution" operator \begin{equation}\label{2.3}
U(t) = < {\rm Tr} e^{iNt H} >.
\end{equation}
We first integrate over the unitary matrix $\omega$ which diagonalizes $H$
in (\ref{2.1});
 without loss of generality we may assume that $H_0$ is a diagonal matrix
with eigenvalues $(a_1, \cdots, a_N)$.
This is done with the help of the well-known Itzykson-Zuber integral for a
unitary matrix
 $\omega$ \cite{Itzykson}, \begin{equation}\label{2.4}
\int
d{\omega} {\rm exp}( {\rm Tr} A {\omega} B {\omega} ^{\dag} ) = {{\rm
det}({\rm exp}(a_i b_j))\over{\Delta(A)\Delta(B)}} \end{equation}
where
$\Delta(A)$ is the Van der Monde determinant constructed with the
eigenvalues of A:
\begin{equation}\label{2.5}
\Delta(A) = \prod_{i<j}^N (a_i - a_j).
\end{equation}
We are then led to
\begin{eqnarray}\label{2.6}
U(t) = && {1\over{Z^{'}\Delta(H_0)}}\sum_{\alpha = 1}^{N} \int
dx_1 \cdots dx_N e^{i Nt x_\alpha} \Delta(x_1,\cdots,x_N) \nonumber\\
&&\times {\rm exp}( - {N\over{2}}\sum x_i^2 + N \sum a_i x_i ).
\end{eqnarray}
The normalization is fixed by
\bq\label{2.7}
U(0) = N
\eq
The integration over the $x_i$'s may be done easily, if we note that
\begin{eqnarray}\label{2.8}
&&\int dx_1 \cdots dx_N \Delta(x_1,\cdots,x_N) {\rm exp}( - {N\over{2}}\sum
x_i^2 + N \sum b_i x_i ) \nonumber\\
&&=
\Delta(b_1,\cdots,b_N) {\rm exp}({{N\over{2}}\sum b_i^2}) \end{eqnarray}

Putting
$
b_i = a_i + i t \delta _{\alpha,i}
$, we obtain
\begin{equation}\label{2.9}
U(t) =\sum_{\alpha=1}^{N} \prod_{\gamma\neq \alpha}^N ({a_\alpha - a_\gamma
+ i t\over{a_\alpha
- a_\gamma}}) e^{-
{Nt^2\over{2}} + i t a_\alpha }
\end{equation}
The sum over N terms in (\ref{2.9})
may then be replaced by a contour-integral in the complex plane,
\begin{equation}\label{2.10}
U(t) = {1\over{i t}} \oint {du\over{2 \pi i}} \prod_{\gamma = 1}^{N} ({u -
a_\gamma + i t\over{ u - a_\gamma}})
e^{- {N t^2\over{2}} + i t N u }
\end{equation}
The contour of integration encloses all the eigenvalues $a_\gamma$. The
Fourier transform
with respect to $t$ gives the density of state
in the presence of an
arbitrary external source $H_0$; note that this representation is exact for
finite $N$.

In the large N limit, the one particle Green function $G(z)$ is readily
obtained \cite{Pastur,BHZ1}.
Indeed one finds that
\begin{eqnarray}\label{2.11}
G(z) &=& < {1\over{N}} {\rm Tr} {1\over{z - {H_{0} + V}}}>\nonumber\\ &=&
{1\over{N}} \sum_{\gamma}
{1\over{z - a_{\gamma} - G(z)}} \end{eqnarray}.
For simplicity we begin with the case in which half of the eigenvalues
$a_{\gamma}$ are equal to
$+a$ and half to $-a$. When $a$ is larger than one, one finds a gap in the
spectrum which
closes when $a$ reaches one.   Then the density of state vanishes at
$\lambda = 0$ from (\ref{2.11}), $\rho(\lambda) \simeq \lambda^{1/3}$. The
density of state is
plotted in Fig.2 of
\cite{BHZ1}.
 
For the n-point correlation function, we have \bq\label{2.12}
R_{n}(\lambda_1,\lambda_2, \cdots, \lambda_n) = <{1\over{N}} {\rm
Tr}\delta(\lambda_1 - H)
{1\over{N}}{\rm Tr}
\delta(\lambda_2 - H ) \cdots{1\over{N}} {\rm Tr}\delta(\lambda_n - H )  >
\eq
By using integral representations for the  $\delta$-functions,
 and repeating the previous steps
we obtain \cite{BH2,BH3}

\bq\label{2.14}
R_2(\lambda_1, \lambda_2) = K_N(\lambda_1, \lambda_1) K_N(\lambda_2,
\lambda_2) -
K_N(\lambda_1, \lambda_2) K_N(\lambda_2, \lambda_1)
\eq
with the kernel
\bq\label{2.15}
K_N(\lambda_1, \lambda_2) = (-1)^{N-1}\int {dt\over{2 \pi}}\oint {du\over{2 \pi
i}}\prod_{\gamma =
 1}^{N} ( {
a_\gamma - i t\over{ u - a_\gamma}}) {1\over{u - i t}} e^{- {N\over{2}}u^2
- {N\over{2}} t^2 - N i
t \lambda_1 + N u \lambda_2}
\eq
Similarly the n-point functions are given in terms of the determinant of
the $n\times n$ matrix
whose elements are $ K_N(\lambda_i,\lambda_j)$ \cite{BH3,Guhr}.
In \cite{BH1}, this kernel $K_N(\lambda_1, \lambda_2)$ was examined in the
large N limit, for fixed
$N(\lambda_1 - \lambda_2)$. In this limit one can evaluate the kernel
(\ref{3.17}) by the
saddle-point method. The result was found to be, up to a phase factor that
we omit here,
\bq\label{2.16} K_N(\lambda_1, \lambda_2) = - {1\over{\pi y}} {\rm sin}
[\pi y \rho(\lambda_1)] \eq
where
$y = N(\lambda_1 - \lambda_2)$.
Apart from the scale dependence provided by the density of state $\rho$,
the n-point correlation
function have thus
a universal scaling limit, i.e. independent of the deterministic part $H_0$
of the
random Hamiltonian.

The large-N behaviour of the density of state $\rho(\lambda) = K(\lambda,
\lambda)$
near the edge
point is also universal. This universality has been investigated for the
Airy case
\cite{Brezin,Tracey} and for the Bessel case \cite{BHZ2}. We will show in
the next section that
similarly at the origin, we find a new class of universality for the
density of states and
for the n-point functions.

\section{CRITICAL BEHAVIOR NEAR THE ORIGIN}
\vskip 5mm
We first consider the case where the eigenvalues of $H_0$ are $\pm a$, each
value being
 N/2 times degenerate. The kernel in (\ref{2.15}) becomes

\bq\label{3.1}
K_N(\lambda_1, \lambda_2) = (-1)^{{N\over{2}} + 1}
\int {dt\over{2 \pi}}\oint {du\over{2 \pi
i}}({a^2 + t^2\over{u^2 -
a^2}})^{{N\over{2}}} {1\over{u - i t}} e^{- {N\over{2}}u^2 - {N\over{2}}
t^2 - N i t \lambda_1 + N u
\lambda_2}
\eq
From this expression,
setting $\lambda_1 = \lambda_2$, we obtain the density of state
$\rho(\lambda_1)$. The derivative
of $\rho(\lambda)$ with respect to
$\lambda $ eliminates the factor $u - i t$ in the denominator of
(\ref{3.1}), and leads to a
factorized expression,

\bq\label{3.2}
{1\over{N}} {\partial\over{\partial \lambda}}\rho(\lambda) = - 
\phi(\lambda)\psi(\lambda)
\eq
where
\bq\label{3.3}
\phi(\lambda) = \int_{-\infty}^{\infty} {dt\over{2\pi}} e^{-{N\over{2}} t^2
+ {N\over{2}}
\ln (a^2 + t^2) - N i t \lambda}
\eq
\bq\label{3.4}
\psi(\lambda) = \oint {du\over{2 \pi i}}
e^{-{N\over{2}}u^2 - {N\over{2}}\ln( a^2 - u^2) + N u \lambda}
\eq
For large $N$ the two integrals defining the functions $\phi$ and $\psi$
are given by a
saddle-point. When
$\lambda_1$ and $\lambda_2$ are near the origin the saddle-points in the
variables $t$ and $u$
move to the origin. Therefore for obtaining the large $N$ behavior of
$\phi$ near $\lambda =0$   we
can expand the logarithmic term in powers of
$t$. One sees readily  that the coefficient of the quadratic term in $t^2$
vanishes for $a = 1$;
in fact three saddle-points are merging at the origin when $a$ reaches one.
We must then expand in
the exponential up to order $t^4$  and we obtain
\bq\label{3.5}
\phi(\lambda) = \int_{-\infty}^{\infty} {dt\over{2 \pi}} e^{- {N\over{4}} t^4 - 
N
i t \lambda}
\eq
Rescaling $t$ to $N^{-1/4} t^{\prime}$ and setting $\lambda =
N^{- 3/4}x$ we find that
\bq\label{3.60}
\hat\phi(x) = N^{1/4}\phi(N^{-3/4}x)
\eq
 has a large $N$, finite $x$,  limit given by

\bq\label{3.6}
\hat\phi(x) = {1\over{\pi}}\int_{0}^{\infty} dt e^{- {1\over{4}} t^4}
\cos ( t x)
\eq
This Fourier transform of the exponential of $-t^4$  is known under the
name of a Turrittin or
Pearcey  integral
\cite{Turrittin,Pearcey}. It is immediate to verify that it satisfies
the differential equation,
\bq\label{3.7}
\hat\phi^{\prime\prime\prime}(x) = x \hat\phi(x) \eq

From the integral representation (\ref{3.6}) we obtain easily the Taylor 
expansion of this function at the origin
\bq\label{3.ya}
\hat\phi(x) = {\sqrt{2}\over{4 \pi}}\sum_{m=0}^{\infty} {\Gamma({1\over{4}}
+{m\over{2}})(-1)^{m} 2^{m} x^{2 m}\over{(2 m)!}} 
\eq
and its asymptotic behavior at large x,
\bq
 \hat \phi(x) \sim \sqrt{{2\over{3 \pi}}} x^{-{1 \over{3}}}
  e^{- {3\over{8}}x^{4\over{3}} }\cos( {3\sqrt{3}\over{8}}x^{4\over{3}} - 
  {\pi\over{6}})
\eq
We return now to the second function (\ref{3.4}).
Similarly, in the scaling limit, N large, 
$\lambda$ small, $N^{3/4} \lambda$ finite, we may expand up to order $u^4$, 
 and define
\bq\label{3.70}
\hat\psi(x) = N^{1/4}\psi(N^{-3/4}x) .
\eq
In the large$N$, finite $x$, limit we find
\bq\label{3.11}
\hat\psi(x) =
\int_c {du\over{2\pi i}} e^{{u^4\over{4}} + u x}.
\eq
This integral is no longer a contour integral around the singularities 
at $u = \pm 1$ ,
but the result of a
saddle-point approximation. Therefore the integral is over a path
consisting of four lines of
steepest descent in the complex u-plane.  Along these
straight lines, the variable
$u$ is changed successively into
$e^{\pm {\pi\over{4}}i} u$ and
$e^{\pm {3
\pi\over{4}} i}u$. This leads to
\bq\label{3.12}
\hat\psi(x) = -{\rm Im} [ {\omega\over{\pi}} \int_{0}^{\infty}
e^{-{u^4\over{4}}}(
e^{x u \omega} - e^{-x u \omega} )] 
\eq
in which $\omega = e^{{\pi i\over{4}}}$. 
The function $\hat\psi(x)$ satisfies the differential equation, 
\bq\label{3.13}
\hat\psi^{\prime\prime\prime}(x) = - x \hat\psi(x) \eq
and again we find from (\ref{3.12}) the Taylor expansion 
\bq\label{3.yb}
\hat\psi(x) = - {1\over{\sqrt{\pi}}} 
\sum_{n=0}^{\infty}{(-1)^n x^{4n + 1} (2n)!\over{n!(4n+1)!}}
\eq
and the large x behavior
\bq
 \hat \psi(x) \sim 2 \sqrt{{2\over{3 \pi}}}x^{-{1 \over{3}}}
  e^{{3\over{8}}x^{4\over{3}} } \cos( {3\sqrt{3}\over{8}}x^{4\over{3}} + 
  {2 \pi\over{3}})
\eq

In fact one may express the whole kernel $K_N(\lambda_1, \lambda_2)$
of (\ref{3.1}) in terms of the two functions $\hat \phi$ and $\hat \psi$ 
in the scaling limit. Indeed defining 
\bq
\lambda_1 = N^{-3/4}x, \lambda_2
=N^{-3/4}y
\eq
\bq\label{3.90}
\hat{K}(x,y) = N^{1/4}K_{N}(N^{3/4}\lambda_1,N^{3/4}\lambda_2) .
\eq
in the large N, finite x and y, limit,
\begin{eqnarray}\label{3.16}
\hat{K}(x,y) &=& \int_{-\infty}^{\infty} {dt\over{2 \pi}}\int_{0}^{\infty}
{du\over{2 \pi
i}} e^{- {t^4\over{4}} - {u^4\over{4}} - i t x} \nonumber\\ &\times&
( {e^{u y \omega}\over{u - i t \omega^{-1}}} - {e^{u y \omega^{-1}}\over{ u
- i t \omega}} + {e^{u
y \omega^{-3}}\over{u - i t \omega^3}} - {e^{u y \omega^3}\over{u - i t
\omega^{- 3}}})
\end{eqnarray} where $\omega = e^{\pi i\over{4}}$
or more explicitely
\begin{eqnarray}\label{3.17}
\hat{K}(x,y) &=& {1\over{2 {\pi}^2 i}} \int_{0}^{\infty} du
\int_{-\infty}^{\infty} dt {e^{- {u^4 + t^4\over{4}} - i t x} \over{u^4 +
t^4}}\nonumber\\
&\times& [ 2 i u^3 \sin \sigma \sinh \sigma + \sqrt{2}u^2 t ( \cos \sigma
\sinh \sigma - \sin \sigma \cosh \sigma)\nonumber\\ &+& 2 i ut^2 \cos
\sigma \cosh \sigma - \sqrt{2} t^3 ( \sin \sigma \cosh \sigma + \cos \sigma
\sinh \sigma )]\nonumber\\ \end{eqnarray}
where $\sigma = y u/\sqrt{2}$.

In Appendix B, it is shown that the kernel $\hat K(x,y)$ may be expressed 
simply in terms of $\hat \phi$ and $\hat \psi$ as
\bq\label{3.x}
\hat K(x,y) = {\hat \phi^{\prime}(x)\hat\psi^{\prime}(y) - 
\hat\phi^{\prime\prime}(x) \hat\psi
(y) - \hat\phi(x)\hat\psi^{\prime\prime}(y)\over{x - y}}
\eq
Therefore the density of state
is given by
\bq\label{3.41}
\hat\rho(x) = - [\hat\phi^{\prime}(x)\hat\psi^{\prime\prime}(x) -
\hat\phi^{\prime\prime}(x) \hat\psi^{\prime}(x)
+ x \hat\phi(x)\hat\psi(x)]
\eq
and the connected correlation function for two eigenvalues
symmetric with respect to the origin
\bq
\hat \rho_c(x,-x) = - [\hat K(x,-x)]^2
\eq
is given through
\bq
 \hat K(x,-x) = {1\over{x}} \phi^{\prime}(x)\psi^{\prime}(x)
\eq
which we know both for x small in an expansion in even powers of x,
 and for large x, at which it behaves as 
\bq
\hat K(x,-x) \sim  {2\over{3 \pi x}} \sin( {3\sqrt{3}\over{4}}x^{4\over{3}} )
\eq

Using Taylor expansions of $\hat \phi(x)$ and $\hat \psi(x)$, 
we have  the expression for 
the kernel $\hat K(x,y)$ by (\ref{3.x}).
In the small $x$ and $y$, it becomes
\begin{eqnarray}\label{3.44}
\hat K(x,y) &=& {1\over{2\sqrt{2}\pi^{3/2}}}[  2 b + {a\over{6}}( - x^2 + 2 x y 
+ 2 y^2)\nonumber\\
 &+& {b\over{60}}( 3 x^4 - 12 x^3 y - 12 x^2 y^2 + 8 x y^3 - 2 y^4)\nonumber\\
 &+& {a\over{5040}}( - 5 x^6 + 30 x^5 y + 30 x^4 y^2 - 40 x^3 y^3 
+ 30 x^2 y^4 - 12 x y^5 \nonumber\\
& & - 12 y^6) 
+ O(x^8)]
\end{eqnarray}
where $a = \Gamma({1\over{4}})= 3.6256$ and $b = \Gamma({3\over{4}})=
1.2254$. In Appendix C, we evaluate the kernel $K_N(x,y)$ for 
several finite N cases, and confirmed (\ref{3.44}).

 The level spacing probability is an important quantity for the universality.
The general distribution is given by the determinant of the kernel 
in our system \cite{BH3}. We consider the probability $E(s)$ of no eigenvalues 
in the 
interval $(- {s\over{2}},{s\over{2}})$.
As same as ordinary random matrix theory without an external source, the 
probability $E(s)$ is given by
\bq\label{3.45}
E(s) = \sum_{n=0}^{\infty} {(-1)^{n}\over{n!}} \int_{-s/2}^{s/2} \cdots
 \int_{-s/2}^{s/2} \prod_{k=1}^{n}dx_k  {\rm det}
 [ K(x_i,x_j)]_{i,j=1,...,n}
\eq
For the small $s$, we have a series expansion for $E(s)$,
\begin{eqnarray}\label{3.46}
E(s) &=& 1 - N\int_{-s/2}^{s/2}\rho(x)dx \nonumber\\
&+& {N^2\over{2}}\int_{-s/2}^{s/2}
\int_{-s/2}^{s/2} [K(x,x)K(y,y) - K(x,y)K(y,x)] dx dy + \cdots
\end{eqnarray}
Using (\ref{3.11}), and $z = N^{-{3\over{4}}}\lambda$, we have
\begin{eqnarray}
E(s) &=& 1 - N^{{3\over{4}}}{\Gamma({3\over{4}})\over{\sqrt{2}\pi^{3/2}}}
\int_{-{s\over{2}}}^{s\over{2}}( 1 + { a\over{4 b}}N^{{3\over{2}}}z^2 
- {1\over{8}}N^3 z^4 + \cdots)dz + \cdots
\nonumber\\
&=& 1 - N^{{3\over{4}}}{\Gamma({3\over{4}})\over{\sqrt{2}\pi^{3/2}}}
(s + {a\over{6 b}}N^{{3\over{2}}}({s\over{2}})^3 - {1\over{20}}
N^3 ({s\over{2}})^5 )+ \cdots
\end{eqnarray}
Then we take a scaling of the energy $s$ as $s\rightarrow N^{-{3\over{4}}}s$,
\bq
E(s) = 1 - {\Gamma({3\over{4}})\over{\sqrt{2}\pi^{3/2}}}(s + {a\over{6 b}}
({s\over{2}})^3 - {1\over{20}}({s\over{2}})^5 )+ \cdots
\eq
Using the expressions of (\ref{3.44}), we have
\bq\label{3.47}
E(s) = 1 - \tilde s + {c^2\over{2}}( {a b\over{9}} s^4 + ({a^2\over{720}}
- {7 b^2\over{900}})s^6 - {41 a b\over{201600}} s^8 ) + O(s^9)
\eq
where $a= \Gamma({1\over{4}}), b = \Gamma({3\over{4}}), c = 1/(2\pi)^{3/2}$ and 
$\tilde s$ is
\begin{eqnarray}\label{3.48}
\tilde s &=& \int_{-s/2}^{s/2} \rho(x) dx \nonumber\\
         &=& 2 c [ b s + {a\over{48}} s^3 - {b\over{20}}({s\over{2}})^5 
         + {a\over{1680}}({s\over{2}})^7 + \cdots ]
\end{eqnarray}
The quantity $\tilde s$ is a function of $s$. We have to use this variable
instead os $s$ since
the density of state is not a constant in the scaling limit, whereas in the
 usual
Wigner-Dyson case, $\tilde s = s$. Therefore, it may be useful to write 
$E(s)$  as a function of $\tilde s$ instead of $s$.
We have in term of $\tilde s$,
\begin{eqnarray}\label{3.49}
E(s) &=& 1 - \tilde s + {a b c^2\over{18}} ({\tilde s\over{2 b c}})^4 - 
({17a^2 c^2\over{4320}} + {7 b^2 c^2\over{1800}})
({\tilde s\over{2 b c}})^6 + O(\tilde s^8)\nonumber\\
&=& 1 - \tilde s + 1.6970 \tilde s^4 - 16.3455 \tilde s^6 + O(\tilde s^8) 
\end{eqnarray}
This result may be  compared with the Wigner-Dyson case, which reads
\bq\label{3.50}
E(s) = 1 - s + {\pi^2\over{36}}s^4 - {\pi^4\over{675}} s^6 + O(s^8)
\eq

The large $\tilde s$ behavior may be obtained through a  Pad\'e analysis of 
the 
series expansion of (\ref{3.49}). In Appendix D, we construct the Pad\'e
 approximants
for the quantity $R(\tilde s)$, defined by
\bq
  R(\tilde s) = {\partial \over{\partial \tilde s}} \ln E(\tilde s)
\eq
and obtain thereby the  large $\tilde s$ behavior of $E(\tilde s)$.

\vskip 5mm
\section{UNIVERSALITY}

We have analyzed the kernel near the origin  when the 
external source eigenvalues are  $a_{\gamma} = \pm 1$. We will
show that the results are independent, in the scaling limit, of 
the distribution of the 
external source eigenvalues (provided a gap closes) and are thus universal.
 We consider, for simplicity, the case in which the external source eigenvalues
$a_{\gamma}$ are distributed symmetrically around the origin.
If we denote the distribution of $a_\gamma$ by $\rho_{0}(a)$, 
the kernel is given by
\begin{eqnarray}\label{4.1}
K(z_1,z_2) &=&- \int_{-\infty}^{\infty}
{dt\over{2\pi}}\oint {du\over{2 \pi i}}{1\over{u - i t}}
e^{- {N\over{2}}t^2 - {N\over{2}}u^2 - N i t z_1 + N u z_2}
\nonumber\\
&\times&
e^{{N\over{2}}\int da \rho_{0}(a) \ln ({a^2 + t^2\over{ a^2 - u^2}})}
\end{eqnarray}
Expanding this expression for small $t$ and $u$, we have
\begin{eqnarray}\label{4.2}
K(z_1,z_2) &=& - \int_{-\infty}^{\infty}
{dt\over{2\pi}}\int_c {du\over{2 \pi i}}{1\over{u - i t}}
e^{- {N\over{2}}t^2 - {N\over{2}}u^2 + {N\over{2}}(t^2 + u^2)
\int da {\rho_{0}(a)\over{a^2}}
- N i t z_1 + N u z_2}
\nonumber\\
&\times&
e^{-{N\over{2}}(t^4 - u^4)\int da {\rho_{0}(a)\over{a^4}}
}
\end{eqnarray}
Thus, if we have
\bq\label{4.3}
\int da {\rho_{0}(a)\over{a^2}} = 1
\eq
we recover  the previous kernel (\ref{3.16}).
The condition of (\ref{4.3}) is  the condition that 
the density of state $\rho(\lambda)$ starts splitting.Indeed
 we consider the equation which determines the  Green function in the large
N limit \cite{BHZ1,Pastur},
\bq\label{4.4}
G(z) = \int da {\rho_{0}(a)\over{z - a - G(z)}} 
\eq
Putting $z = 0$, noting that parity implies that Re $G(0) = 0$,
when we take the imaginary part of (\ref{4.4}), we 
obtain
\bq\label{4.5}
\rho(0) = \rho(0) \int da {\rho_{0}(a)\over{a^2 + [\pi\rho(0)]^2}}
\eq
Therefore, as long as $\int da {\rho_{0}(a)\over{a^2}}<1$  we find that
the condition of (\ref{4.3}) is equivalent to $\rho(0) = 0$, a gap is present, 
but when
 the condition of (\ref{4.3}) is satisfied the gap closes and we recover the 
kernel
that we have considered hereabove.
Therefore in  the scaling regime $K(z_1,
z_2)$ near $z_1 = z_2
 = 0$ is indeed universal, i.e. independent 
of $\rho_{0}(a)$ provided the condition (\ref{4.3}) is satisfied, or 
equivalently that a gap closes at the origin in the density of states.

\vskip 5mm
\section{MULTI-CRITICAL BEHAVIOR}
We have obtained the universal behavior of the kernel near 
the origin by the scaling of the energy $\lambda = N^{ -3/4}
x$ in the large N limit with a fixed $ x$,
when the spectrum  of $H_0$ is such that 
 the quadratic terms in  $t$ in the exponent of the 
integrand   vanish.
Now we discuss the multi-critical behavior in which the exponent
starts from  $t^{2k + 2}$ term in the scaling limit, and the energy $\lambda$
is scaled by $\lambda = N^{ -{2k + 1\over{2k + 2}}} x$.

 Suppose that we begin with  the Hermitian matrix $H$ coupled to 
 $H_{0}$, which is 
 a diagonal complex matrix. 
 The probability distribution of the matrix $H$ is 
\bq\label{5.1}
   P(H) = {1\over{Z}} e^{ - {N\over{2}} {\rm Tr} ( H^2 - 2 H_{0} H )}
\eq
 Then, this probability becomes complex, and we loose the meaning 
 of a  probability distribution since it is not positive definite. We here 
 simply pursue the analytic continuation of the 
 diagonal eigenvalues $a_\gamma$ in the complex plane.
 By choosing the appropriate $a_{\gamma}$, we can  obtain an  effective
kernel, which starts with  $t^{2 k + 2}$ term in the exponent of the integrand.

\bq\label{5.2}
K(z_1,z_2) = - \int {dt\over{2 \pi}} \oint {du\over{2 \pi i}}
{1\over{u - i t}} e^{ - N C ( t^{2 k + 2} + (-1)^k 
u^{2 k + 2} ) - N i t z_1 + N u z_2}
\eq

For $k = 2$, the eigenvalues of the  external source  $a_{\gamma}$ are
 assumed to take the values $a_{\gamma} = \pm a, \pm b$, each one beeing 
${N\over{4}}$
times degenerate.
For higher  multi-criticality, the conditions are 
\bq\label{5.3}
   {1\over{a^2}} + {1\over{b^2}} = 2
\eq
\bq\label{5.4}
   {1\over{a^4}} + {1\over{b^4}} = 0,
\eq
We then  get an  effective action for the kernel which starts with a $t^6$ term
 (\ref{5.2}) in the large N limit. The solution of (\ref{5.3}) and (\ref{5.4})
are $a = {1\over{2^{1/4}}} e^{- {\pi\over{8}}i}$, and 
$b = {1\over{2^{1/4}}} e^{{\pi\over{8}}i}$. The effective form of 
the kernel is thus
\bq\label{5.5}
K(\lambda,\mu) = - \int_{-\infty}^{\infty} {dt\over{2 \pi}}
\oint {du\over{2 \pi i}}{1\over{u - i t}}e^{-{N\over{3}}(t^6 + 
u^6) - N i t \lambda + N u \mu}
\eq
The scaling limit is obtained by rescaling  $t$ and $u$  by $N^{-{1\over{6}}}$ 
and
$\lambda$ and $\mu$  by $N^{-{5\over{6}}}$.
The integration of (\ref{5.5}) shows a Painlev\'e II equation of $A_{4}$ type.

Similarly to (\ref{3.x}), we find
\begin{eqnarray}\label{5.9}
\hat K(x,y) &=& {2 \over{x - y}}[ \phi^{\prime\prime\prime\prime}(x)\psi(y) - 
\phi^{\prime\prime\prime}(x)\psi^{\prime}(y) +
\phi^{\prime\prime}(x)\psi^{\prime\prime}(y) -
\phi^{\prime}(x)\psi^{\prime\prime\prime}(y) \nonumber\\
& & + \phi(x)\psi^{\prime\prime\prime\prime}(y) ]
\end{eqnarray}
where $\phi(x)$ and $\psi(y)$ are defined by
\bq
 \phi(x) = \int_{-\infty}^{\infty} {dt\over{2 \pi}} e^{-{t^6\over{3}} + i t x}
\eq
\bq
\psi(x) = \int_c {du\over{2 \pi i}} e^{-{u^6\over{3}} + u x}
\eq
The path $c$ of the integration is taken along four
lines, starting from the origin to infinity, and $u$ is replaced on this lines
as $u =  e^{\pm{\pi\over{3}}i}u^{\prime}$, 
and $u = e^{\pm
{2 \pi\over{3}}i}u^{\prime}$. These functions satisfy the differential 
equations,
\bq
{\rm {d^5\over{d x^5}}} \phi(x) = - {1\over{2}} x \phi(x)
\eq
\bq
{\rm {d^5\over{d x^5}}} \psi(x) =  {1\over{2}} x \psi(x)
\eq


In general, if we choose $a_{\gamma}= (\pm a_1,...,\pm a_k)$, each $a_j$ is 
$N/2 k$ times degenerated, and if they satisfy
\begin{eqnarray}\label{5.14}
\sum_{i = 1}^k {1\over{a_i^2}} &=& k\nonumber\\
\sum_{i = 1}^k {1\over{a_i^{2m}}} &=& 0, (m = 2, 3, ..., k)
\end{eqnarray}
the effective kernel becomes (\ref{5.2}).
The scaling $t = N^{- {1\over{2 k + 2}}} t^{\prime}$, and 
$z = N^{-{2 k + 1\over{2 k + 2}}}\lambda$ 
gives the universal behavior in the large N limit.
\vskip 5mm

\section{DEGENERACY OF BESSEL CASE }
  In a previous article \cite{BHZ2}, we have considered an ensemble
of matrices $M$,of the form
 \bq\label{6.1}
 M = \left ( \matrix{0 &C^{\dag} \cr
             C &0 \cr }\right )
 \eq
 in which $C$ is an $N \times N$ complex random matrix, with 
 probability distribution,
 \bq\label{6.2}
 P(C) = {1\over{Z}} {\rm exp}( - N {\rm Tr} C^{\dag} C - N {\rm Tr} A C^{\dag} C 
)
 \eq
 Noting that
\begin{eqnarray}\label{6.3}
G(z) &=& <{1\over{2 N}} {\rm Tr} {1\over{z - M}} >\nonumber\\
&=& < {1\over{N}} {\rm Tr} {z\over{z^2 - C^{\dag} C}}>
\end{eqnarray}
we obtain a  relation between the density of states,
\bq\label{6.4}
\rho(z) = <{1\over{2N}} {\rm Tr} \delta( z - M ) >
\eq
and
 \bq\label{6.5}
 \tilde\rho(z) = < {1\over{N}} {\rm Tr} \delta( z - C^{\dag} C ) >
 \eq
which reads
\bq\label{6.6}
\rho(z) = | z| \tilde\rho(z^2)
\eq
 The two point correlation function is 
 \bq\label{6.7}
 R_2(z_1,z_2) = < {1\over{N}} {\rm Tr} 
 \delta( z_1 - C^{\dag} C ) {1\over{N}} {\rm Tr} \delta( z_2 - C^{\dag} C )>
 \eq
 Using the Itzykson-Zuber formula, we obtain the kernel in the presence of the 
 external source $A$, 
 \bq\label{6.8}
 K(z_1,z_2) = \int_{-\infty}^{\infty} {dt\over{2 \pi}}
 \oint {du\over{2 \pi i}} ( {1\over{u - i t}} )({1 + u\over{1 + i t}})^N
 \prod_{\gamma} {( a_\gamma - i t)\over{( u - a_\gamma)}} 
 e^{- i t N z_1 + u N z_2}
 \eq
 When we put $a_\gamma = 0$ and take the large N limit, with
shifts $t,u\rightarrow Nt, N u$
  we obtain the Bessel kernel\cite{BHZ2}.
 
 If we choose $a_\gamma = - 1 + a, -1 + b$, each one is $N/2$ times
 degenerated, and  making the shifts $u\rightarrow u - 1$ and $t \rightarrow
t + i$,
 the kernel is written by
 \begin{eqnarray}\label{6.9}
 K(z_1,z_2) &=& \int_{-\infty}^{\infty} {dt\over{2 \pi }} \oint {du\over{2 \pi 
i}}
 {1\over{u - i t}}({u\over{i t}})^N [{( a - i t) ( b - i t)\over{( u - a) ( u - 
b)}}]^{
 {N\over{2}}}\nonumber\\
&\times& e^{- i Nt z_1 + N u z_2 + N(z_1 - z_2)}\nonumber\\
 &=&\int_{-\infty}^{\infty} {dt\over{2 \pi }} \oint {du\over{2 \pi i}}
 {1\over{u - i t}} e^{{N\over{2}}\ln ( 1 - {a\over{i t}}) + {N\over{2}}
 \ln ( 1 - {b\over{i t}}) - {N\over{2}} \ln ( 1 - {a\over{u}}) - {N\over{2}}
 \ln ( 1 - {b \over{u}})}\nonumber\\
 &\times&
  e^{- i Nt z_1 + N u z_2 + N(z_1 - z_2)}
 \end{eqnarray}
 Taking the large N limit with  
scalings $t \rightarrow N t$ and $u \rightarrow N u$,
we have after expanding the logarithm to first order,
\begin{eqnarray}\label{6.10}
K(z_1,z_2) &=& N \int_{-\infty}^{\infty} {dt\over{2 \pi}}
\oint {du\over{2 \pi i}} {1\over{u - i t}}
e^{{ i\over{2 t}} (a + b) + {1\over{2 u}} (a + b)}\nonumber\\
&\times& e^{ - i N^2 t z_1 + N^2 u z_2 + N ( z_1 - z_2)}
\end{eqnarray}
If $a + b$ is nonvanishing, we have a Bessel kernel,
 which we have studied in an earlier publication.
\cite{BHZ2}. The density of state
is obtained as  the derivative of $K(z,z)$ ; for example, in the case 
$a + b = 2$
\bq\label{6.11}
 {\partial K(z,z)\over{\partial z}} = N^3 \int {dt\over{2 \pi}}
e^{{i\over{t}} - i t x^2} \oint {du\over{2 \pi i}}
e^{{1\over{u}} + u x^2}
\eq
where we have put $z = N^{-2} x^2$. The right hand side of (\ref{6.11}) becomes
$N^3J_1^2(2 x)/4 x^2$, and the density of state
in (\ref{6.6}) has the universal form,
\bq\label{6.12}
\rho(z) = N z [ J_0^2(2 z N) + J_1^2( 2 z N)]
\eq

For the Bessel kernel of (\ref{6.10}), we  have an integral representation.
 By an  appropriate rescaling, the Bessel kernel  reduces to \cite{Tracey}
\bq\label{6.12a}
K(x,y) = \int_{0}^{\infty} \phi(x + z) \phi(y + z) dz
\eq
where
\bq\label{6.12b}
  \phi(x) = {J_{1}(\sqrt{x})\over{\sqrt{x}}}
\eq
Using this representation, and the standard relations between Bessel 
functions $J_{0}(x)$ and $J_{1}(x)$, and by considering
\bq\label{6.12c}
I_{1}= \int_{0}^{\infty} ( x + z) {J_{1}(\sqrt{x + z})\over{\sqrt{x + z}}}
{J_{1}(\sqrt{y + z})\over{\sqrt{y + z}}} dz
\eq
we easily obtain 
\begin{eqnarray}\label{6.12d}
I_1 - I_2 &=& ( x - y)K(x,y)\nonumber\\
&=& \sqrt{x}J_1(\sqrt{x})J_0(\sqrt{y}) - \sqrt{y}J_1(\sqrt{y})J_0(\sqrt{x})
\end{eqnarray}
where $I_2 = I_1( x\rightarrow y, y\rightarrow x)$.
We have thus obtained  a closed expression for the kernel from this equation;
we could have obtained it 
also through the Christoffel-Darboux identity for  orthogonal
polynomials( Laguerre polynomials in this case).

When $a + b = 0$, we  find a new  scaling limit near the 
origin. Expanding the logarithmic terms in (\ref{6.9}), we obtain
for the kernel, after  rescaling $z_1 = N^{-{3\over{2}}}
x^2$, $t = N^{-{1\over{2}}}t^{\prime}
$,
\bq\label{6.13}
I = \int_{-\infty}^{\infty}
{dt^{\prime}\over{2 \pi}} {\rm exp} [ - {1\over{4 t^2}}(a^2 + b^2)
- i t x^2]
\eq
This integral should be understood as a  contour integral and it should be
calculated by taking the residue at the origin. 
This integral $I$ appears in Painlev\'e type III with $A_2$ system,
two-dimensional Garnier system \cite{Okamoto}.  
The perturbative expansion of this kernel may be deduced from this  
integral representation.

The higher multi-critical points are obtained by tuning the values 
of $a_\gamma$; however the next one, for instance, requires the 
conditions; $a_1 + a_2 + a_3 + a_4 = 0$, $a_1^2 + a_2^2 + a_3^2 + a_4^2
= 0$ $(a_\gamma = a_1,a_2,a_3,a_4)$. Thus, it is necessary to consider
the complex values for $a_\gamma$, for example, $a_1 = 1,a_2 = -1,a_3 = i,
a_4 = -i$. 
\vskip 5mm

\section{SUMMARY AND DISCUSSIONS}

In this paper, we have considered the critical scaling 
  behavior when the two edges of the support of the eigenvalues merge 
  at one point. This has been studied within  
  the Gaussian random matrix model with an external source.
 The correlation functions and the level spacing distribution belong then to 
a new universality class, differnet from the Wigner-Dyson case. 
  
    Although our model is limited to  Gaussian potentials,
 several  universality
    classes may  be obtained by tuning the external source eigenvalues.
    It may be interesting to extend the study to more complicated
    cases like two dimensional gravity, for which the kernel has
    a similar expression  \cite{Tracy2}.
  
 The unitary matrix model at the transition point 
has also  been shown to be related to  the Painlev\'e II type
by considering  non-Gaussian distributions  on the unit circle ;
multi-critical behavior has also been investigated there\cite{Periwal}. 
Our Gaussian model
with an external source is an  alternative model 
 for tuning multi-critical
behavior; it gives  concise and exact expressions for finite $N$, which
are useful to derive simple closed expressions for the 
universal correlation functions in the scaling limit.

 We have considered the critical case when a gap opens in the density
 of states.
Even in this case, the n-level correlation functions are determined by the 
kernel, completely. Therefore, we know that all the 
 correlation functions become
universal. After opening the gap, there is no more universality. Indeed, 
if we average over the external source eigenvalues in some interval, 
when the gap opens,
we obtain a Poisson like behavior for the level
 spacing probability distribution.

 It has been shown that the Bessel kernel case has interesting
physical applications in several different problems, including
the random flux problem, two-state quantum Hall effect \cite{BHZ2,HSW},
and the zero mode of a Dirac operator \cite{Verbaarschot}.
It should  be interesting to discuss the physical applications  of 
the multi-critical behavior  discussed in this paper.
\acknowledgements
 This work was supported by the CREST of JST. 
\vskip 5mm
\appendix

\section{Two dimensional Painlev\'e II equation }
\vskip 5mm
It is known that the differential equation (\ref{3.8}) 
is related to a Painlev\'e II
differential equation
with an $A_2$ system, where the notation $A_2$ stands for a two-dimensional
Garnier system
\cite{Garnier,Okamoto}. 
 When two saddle points merge, instead of three, the exponent in
(\ref{3.5}) is only $t^3$,
and its Fourier transform is an Airy function; it
is then related to Painlev\'e II with an  $A_1$ system,
 and the singurality near the edge of the density of state is
indeed described in terms
of Airy functions
\cite{Brezin,Tracey}. In our case, two edges merge at the origin, three
saddle-points collapse
together, and the Airy case degenerates.

Another mathematical interpretation of these functions may be obtained as
follows. Let us note
that the Painlev\'e II equation,  is given by the non-linear equation
\bq\label{3.8}
{d^2 q\over{dt^2}} = 2 q^3 + t q + \alpha \eq
This equation is equivalent to the two equations,
\bq\label{3.9}
{d q\over{d t}} = p - q^2 - {t\over{2}}
\eq
\bq\label{3.10}
{d p\over{d t}} = 2 p q + \alpha + {1\over{2}} \eq
These two equations in turn, may be viewed as Hamilton's equations based on
$H(p,q) = {1\over{2}}p^2
- p( q^2 + {1\over{2}}t) - (\alpha + {1\over{2}})q $. Taking $\alpha=
{-1/2}$, $p=0$ is a solution;
writing then  $q = {d\over{dt}} {\rm ln} \phi$, one finds that $\phi$
satisfies $[\phi(t)]''=
-{t\over{2}}\phi(t)$  and thus
$\phi$ is an Airy function.  

The two-dimensional Hamiltonians for the Painlev\'e II equation ($A_2$ system) 
are \cite{Okamoto}
\begin{eqnarray}
  H_1 &=& (q_2^2 - q_1 - t_1)p_1^2 + 2 q_2 p_1p_2 + p_2^2 + 
2 (q_1^2 + t_2 q_2 - t_1^2) p_1 \nonumber\\
&+& 2 (q_1 q_2 + t_1 q_2 + t_2) p_2 + \kappa q_1
\end{eqnarray}
\bq
H_2 = q_2 p_1^2 + 2 p_1 p_2 + 2 (q_1 q_2 + t_1 q_2 + t_2) p_1 
+ 2( q_2^2 - q_1 + t_1) p_2 + \kappa q_2
\eq
We have, (i, j = 1, 2)
\bq
  {\partial q_j\over{\partial t_i}} = {\partial H_i\over{\partial
p_j}}, {\partial p_j\over{\partial t_i}} = -{\partial H_i\over{\partial
q_j}} 
\eq
When $\kappa = 0$, the particular solution is given by 
\begin{eqnarray}
q_j(t) &=& - {1\over{2}}{\partial\over{\partial t_j}}
\ln [ e^{-t_1^2}w(t_1,t_2)]\nonumber\\
p_j(t) &=& 0
\end{eqnarray}
Then, $w(t_1,t_2)$ satisfies the linear partial differential equations,
\begin{eqnarray}
  {\partial^2 w\over{\partial t_1^2}} &=& 4 t_1 {\partial w\over{
\partial t_1}} + 2 t_2 {\partial w\over{\partial t_2}} + 2 w
\nonumber\\
{\partial^2 w\over{\partial t_1 \partial t_2}} &=& 4 t_1 {\partial w
\over{\partial t_2}} - 4 t_2 w \nonumber\\
{\partial^2 w\over{\partial t_2^2}} &=& - 2 {\partial w\over{\partial
t_1}}
\end{eqnarray}
If we put $t_1 = 0$ in these equations, we obtain
\bq\label{AA}
  {\partial^3 w\over{\partial t_2^3}} = 8 t_2 w
\eq
In general, $w$ is given by
\bq
  w = \int_c e^{- {u^4\over{2}} - 2 t_1 u^2 - 2 t_2 u} du
\eq
The function $\hat \phi(x)$ in (\ref{3.7}) does 
satisfy  the same differential equation
(\ref{AA}).

\vskip 8mm
\section{ The kernel $\hat K(x,y) $ 
 }
\vskip 5mm

We have the following equation for
$K(x,y)$, which is derived by the contour integral representation
(\ref{3.1}),
\bq\label{3.34}
{\partial\over{\partial z}}\hat K(x + z, y + z) = - \hat \phi(x + z)
\hat \psi(y + z)\eq
From this equation, we have
\begin{eqnarray}
 & &(x - y) {\partial\over{\partial z}}\hat K(x + z, y + z) =
  - [(x + z) - (y + z)]
\hat \phi(x + z)
\hat \psi(y + z)\nonumber\\
&=& - (\hat \phi^{\prime\prime\prime}(x + z) \hat \psi(y + z) + 
\hat \phi(x + z) \hat \psi^{\prime\prime\prime}(y + z) )\nonumber\\
&=& - {\partial\over{\partial z}}
[ \hat \phi^{\prime\prime}(x + z) \hat \psi(y + z) + 
\hat \phi(x + z) \hat \psi^{\prime\prime}(y + z) - 
\hat \phi^{\prime}(x + z) \hat \psi^{\prime}(y + z) ]\nonumber\\
\end{eqnarray}
Therefore, we get by integration,
\begin{eqnarray}\label{3.z2}
& &(x - y)\hat K(x + z, y + z)\nonumber\\
& &= - [  \hat \phi^{\prime\prime}(x + z) \hat \psi(y + z) + 
\hat \phi(x + z) \hat \psi^{\prime\prime}(y + z) - 
\hat \phi^{\prime}(x + z) \hat \psi^{\prime}(y + z) ] \nonumber\\
& & + C(x,y)
\end{eqnarray}
Setting $z = 0$ in this equation, it is then easy to prove
\bq\label{3.z1}
({\partial\over{\partial x}} + {\partial\over{\partial y}}) C(x,y) = 0
\eq
since we have
\bq
({\partial\over{\partial x}} + {\partial\over{\partial y}})\hat K(x,y) = -
\hat\phi(x)\hat\psi(y)
\eq
Thus, we find from (\ref{3.z1}) $C(x,y) = C(x - y)$. 
Setting further $y = 0$ in (\ref{3.z2}), and noting that $\psi(0) =
\psi^{\prime\prime}(0) = 0$ and  from (\ref{3.16})
\bq
 x \hat K(x, 0) = \hat \phi^{\prime}(x) \hat \psi^{\prime}(0)
\eq
we find  that the integral constant $C(x)$ is vanishing.

Thus, we have a simple expression for $\hat K(x,y)$ by
\bq\label{3.xx}
\hat K(x,y) = {\hat \phi^{\prime}(x)\hat\psi^{\prime}(y) - 
\hat\phi^{\prime\prime}(x) \hat\psi
(y) - \hat\phi(x)\hat\psi^{\prime\prime}(y)\over{x - y}}
\eq
which is same as (\ref{3.x}).

Since $\hat K(x,y)$ should be finite for $y\rightarrow x$, we have the following
constraint between $\hat\phi(x)$ and $\hat\psi(x)$, 
\bq\label{3.40}
\hat\phi^{\prime}(x)\hat\psi^{\prime}(x) - \hat\phi^{\prime\prime}(x)\hat\psi(x) 
-
\hat\phi(x)\hat\psi^{\prime\prime}(x) = 0
\eq
This identity is proved by considering the derivative,
\begin{eqnarray}
& &{d\over{d x}}[\hat \phi^{\prime}(x)\hat\psi^{\prime}(x) - 
\hat\phi^{\prime\prime}(x) \hat\psi
(x) - \hat\phi(x)\hat\psi^{\prime\prime}(x)] \nonumber\\
& & = - \hat \phi^{\prime\prime\prime}(x)\hat\psi(x) - 
\hat \phi(x)\hat\psi^{\prime\prime\prime}(x)
\nonumber\\
& &= 0
\end{eqnarray}
Then, by integration, we find the l.h.s of (\ref{3.40}) is a constant, but 
it vanishes since, by putting $x = 0$, we have $\hat \phi^{\prime}(0) = 
\hat \psi(0)
= \hat \psi^{\prime\prime}(0) = 0$.
from (\ref{3.ya}) and (\ref{3.yb}).

\vskip 8mm

\section{ Finite N calculation 
 }
\vskip 5mm
We consider some finite N exact calcualtions for the density of state
and for the correlation function near $\lambda = 0$.

The derivative of the density of state is factorized as (\ref{3.2}).
We evaluate $\phi_N(\lambda)$ of (\ref{3.3}).
The expansion at small $\lambda$ is easily done, and after the rescaling 
$x = N^{-3/4}\lambda$, it gives, for instance in the case $N=6$,
\bq\label{A.1}
\phi(x) = \phi(0) ( 1 - 0.47525 
x^2 + 0.09017 x^4 + \cdots )
\eq
which should be compared with the expansion
\ref{3.ya})in the large $N$ limit
 $\phi(x) = \phi(0) ( 1 - 0.33798 
x^2 + 0.04166 x^4 + \cdots )$.
For $\psi(\lambda)$, we determine the expansion for small $\lambda$
 by computing  the residue in (\ref{3.4}) as,
\begin{eqnarray}\label{A.2}
\psi(\lambda) &=& C_N N^{-{1\over{2}}}( - \lambda - {1\over{3!}}({2\over{N}})
\lambda^3 + 
{1\over{5!}}({2\over{N}} - {8\over{N^2}}) \lambda^5 \nonumber\\
&+& {1\over{7!}}( {20\over{N^2}} - {48\over{N^3}}) \lambda^7 + \cdots)
\end{eqnarray}
where $\lambda = N^{{1\over{4}}} x$ and $C_N$ is a coefficient, 
which tends to $1/\sqrt{\pi}$. Thus, in the large N limit, for a 
fixed $x$, we get
\bq\label{A.3}
\hat \psi(x) = - {1\over{\sqrt{\pi}}}  x ( 1 - {2\over{5!}} x^4
+ {12\over{9!}} x^8 \cdots )
\eq
There are $x^3$ and $x^7$ term, but they are order of $N^{-{3\over{
4}}}$ and neglected in the large N limit. This result agrees with 
(\ref{3.yb}) completely.

$K_N(x,y)$ is evaluated for small N (N = 2,4,6) by the contour integration 
(\ref{3.1}) to confirm
the universal form of (\ref{3.44}).
We have
\bq\label{A.5}
K_2(x,y) = {\sqrt{\pi}\over{2 \pi}} e^{-1}[ 1 + ( 2 x y + 2 y^2 - x^2) 
+ ({2\over{3}} y^4 + {4\over{3}} x y^3 - 2 x^2 y^2 - 2 x^3 y + {1\over{2}}
x^4) + \cdots]
\eq
\begin{eqnarray}\label{A.6}
K_4(x,y) &=&  {1\over{2 \pi}}\sqrt{{\pi\over{2}}} e^{-2}
[ {7\over{2}} + 9 ( 2 x y + 2 y^2 - x^2)\nonumber\\
 &+& 22 ({32\over{66}}y^4 + 
{4\over{3}} x y^3 - 2 x^2 y^2 - 2 x^3 y + {1\over{2}} y^4) + \cdots ]
\end{eqnarray}
\begin{eqnarray}\label{A.7}
K_6(x,y) &=& {1\over{2 \pi}} \sqrt{{\pi\over{3}}} e^{-3}
[ {87\over{8}} + {393\over{8}} ( 2 x y + 2 y^2 - x^2)\nonumber\\
&+& {1647\over{8}}({630\over{1647}} y^4 + {4\over{3}} x y^3 - 2 x^2 y^2 
- 2 x^3 y + {1\over{2}} x^4) + \cdots ]
\end{eqnarray}
where we have to scale $x$ by $N^{-{3\over{4}}} x$ to compare with (\ref{3.44}).
After the scaling, we find that it tends to the universal form in the 
large N limit.


\section{Pad\'e Approximation for the large 
 $\tilde s$ }
\vskip 5mm

For the Wigner-Dyson case, the probability of no eigenvalue in the interval 
$(-{s\over{2}},
{s\over{2}})$, E(s) is Gaussian for large $s$. Therefore the quantity $R(s)$ 
defined by
\bq
  R(s) = {\partial\over{\partial s}} \ln E(s)
\eq
is proportional to $s$ in the large N limit.

We apply Pad\'e analysis for $R(s)$ in the Wigner-Dyson case, in which the exact 
expression for $R(s)$ is known. Using the series of $E(s)$ in (\ref{3.50}), we 
have
\begin{eqnarray}
R(s) &=& - [ 1 + s + s^2 + ( 1 - {\pi^2\over{9}}) s^3 + 
(1 - {5 \over{36}} \pi^2) s^4
 + ( 1 - {\pi^2\over{6}} + {2\pi^4\over{225}}) s^5 \nonumber\\
&+&   ( 1 - {7\pi^2\over{36}} + {7\pi^4\over{
675}}) s^6 + 
(1 - {2\pi^2\over{9}} + {121 \pi^4\over{8100}} - {\pi^6\over{2205}})s^7
+ ...]
\end{eqnarray}
The [3,2] Pad\'e approximant for $R(s)$ is given by
\bq
  R(s) = - ({1 + a_1 s + a_2 s^2 + a_3 s^3 \over{1 + b_1 s + b_2 s^2}})
\eq
where $a_1 = 1.486956,a_2 = 1.90478, a_3 = 0.808162, b_1 = 0.486956, b_2 = 
0.417829$. Thus,
in the large $s$ limit, $E(s)$ is estimated to  exp( $- 0.967097 s^2$). 
The exact value is known to be 
$E(s) \sim$ exp($ - \pi^2 s^2/8$), where $\pi^2/8 = 1.233$.
With the [4,3] Pad\'e, the estimate is further improved as 
$a_1 = 1.85662,a_2 = 2.56953, a_3 = 1.63848, 
a_4 = 0.424931, b_1 = 0.856623, b_2 = 0.712904,
b_3 = 0.165575$, and $E(s) \sim $ exp($- a_4 s^2/2 b_3 $), where 
$a_4/2 b_3 = 1.2832$. This value is very close to the exact solution.
Using this Pad\'e approximation for $R(s)$, we obtain the behavior of
$E(s)$ for all region of $s$  by
\bq
    E(s) = \exp[ \int_0^s dx R(x)]
\eq
    The level spacing probability $p(s)$ is obtained as 
the second derivative of $E(s)$.
\begin{eqnarray}\label{ps}
    p(s) &=& {d^2\over{d s^2}} E(s)\nonumber\\
    &=& [ {d\over{d s}} R(s) + (R(s))^2] \exp[ \int_0^s dx R(x)]
 \end{eqnarray}
 The value of $p(s)$ by [4,3] Pad\'e analysis for $R(s)$ is shown in Fig.1.
 
 When we apply this Pad\'e analysis for (\ref{3.49}) in terms of $\tilde s$, 
we have for [3,2] Pad\'e,
$a_1 = 2.1118,a_2 = 16.0318, a_3 = 9.24379, b_1 = 1.11179, b_2 = 13.920$, and 
$E(\tilde s) \sim$ exp($ - 0.3320 \tilde s^2$). In the large $\tilde s$ limit,
$\tilde s$ is proportional to $s^{4/3}$ from (\ref{3.48}).
 The level spacing probability $p(s)$ in this case is obtained from the second
 derivatives of $E(\tilde s)$ by $\tilde s$. The result of  [3,2] Pad\'e for 
 $p(\tilde s)$  is shown in Fig.1.


\end{document}